\documentclass{article}
\usepackage{spconf,amsmath,graphicx,amsfonts}
\usepackage{booktabs,multirow,bigdelim}

\usepackage{pifont}
\newcommand{\cmark}{\ding{51}}%
\newcommand{\xmark}{\ding{55}}%
\usepackage{enumitem}
\usepackage[labelformat=simple]{subcaption}

\usepackage[font=small,labelfont=bf]{caption}
\usepackage{adjustbox}
\usepackage{graphicx}
\usepackage{tikz}
\usepackage{flushend}
\usepackage{hyperref}
\allowdisplaybreaks

\usepackage{math/math_defs}
\usepackage{bbm}

\def\rms{\mathrm{s}}
\def\rmn{\mathrm{n}}
\def\musi{\muvec_{\rms_i}}
\def\muni{\muvec_{\rmn_i}}
\def\mus{\muvec_{\rms}}
\def\mun{\muvec_{\rmn}}
\def\Ls{\Lambmat_{\rms}}
\def\Ln{\Lambmat_{\rmn}}
\def\Lsn{\Lambmat_{\rms\rmn}}

\def\Lss{\Lambmat_{\rms\rms}}

\def\Ssn{\Sigmat_{\rms\rmn}}
\def\Sns{\Sigmat_{\rmn\rms}}
\def\Sss{\Sigmat_{\rms\rms}}
\def\Snn{\Sigmat_{\rmn\rmn}}

\def\rsi{r_{\rms_i}}
\def\rni{r_{\rmn_i}}
\def\Nsi{N_{\rms_i}}
\def\Nni{N_{\rmn_i}}
\def\Fsi{\Fvec_{\rms_i}}
\def\Fni{\Fvec_{\rmn_i}}
\def\Ssi{\Smat_{\rms_i}}
\def\Sni{\Smat_{\rmn_i}}
\def\rs{r_{\rms}}
\def\rn{r_{\rmn}}

\usetikzlibrary{decorations.pathmorphing}
\newcommand{\cutimage}[3]{
  \adjustbox{max height=2in, max width=\linewidth}{
    \tikz{
      \node[inner sep=0pt] (A) {
        \adjustbox{trim={0} {#2\height} {0} {0},clip}{\includegraphics[width=\linewidth]{#1}}
      };
      \draw[decoration={zigzag, mirror,segment length=2mm,amplitude=0.5pt}, decorate](A.south west) -- (A.south east)
    }
  }
  \adjustbox{max height=2in, max width=\linewidth}{
    \tikz{
      \node[inner sep=0pt] (A) {
        \adjustbox{trim={0} {0} {0} {#3\height},clip}{\includegraphics[width=\linewidth]{#1}}
      };
      \draw[decoration={zigzag, mirror,segment length=2mm,amplitude=0.5pt}, decorate](A.north west) -- (A.north east)
    }
  }
}

\title{Frustratingly easy noise-aware training of acoustic models}
\name{Desh Raj$^1$, Jes\'us Villalba$^{1,2}$, Daniel Povey$^3$, Sanjeev Khudanpur$^{1,2}$\thanks{This work was partially supported by grants from the JHU Applied Physics Laboratory, MIT Lincoln Labs, and the Government of Israel.}}
\address{
  $^1$Center for Language and Speech Processing \& $^2$Human Language Technology Center of Excellence \\ 
  The Johns Hopkins University, Baltimore, MD 21218, USA.\\
  $^3$Xiaomi Corp., Beijing, China.}
\email{\small{\texttt{draj@cs.jhu.edu}}}

\begin{document}

\maketitle

\begin{abstract}
Environmental noises and reverberation have a detrimental effect on the performance of automatic speech recognition (ASR) systems. Multi-condition training of neural network-based acoustic models is used to deal with this problem, but it requires many-folds data augmentation, resulting in increased training time. In this paper, we propose utterance-level noise vectors for noise-aware training of acoustic models in hybrid ASR. Our noise vectors are obtained by combining the means of speech frames and silence frames in the utterance, where the speech/silence labels may be obtained from a GMM-HMM model trained for ASR alignments, such that no extra computation is required beyond averaging of feature vectors. We show through experiments on AMI and Aurora-4 that this simple adaptation technique can result in 6-7\% relative WER improvement. We implement several embedding-based adaptation baselines proposed in literature, and show that our method outperforms them on both the datasets. Finally, we extend our method to the online ASR setting by using frame-level maximum likelihood for the mean estimation.  

\end{abstract}
\noindent\textbf{Index Terms}: robust speech recognition, noise-aware training, acoustic modeling, noise vectors

\section{Introduction}

Automatic speech recognition (ASR) systems are being increasingly deployed in real-life situations (e.g. home assistants like Amazon Echo, Google Home, etc.), where the acoustic environment may present distortions such as background noise and reverberation. To tackle these problems, the field of noise-robust ASR has organized various challenges over the past several years, such as the REVERB~\cite{kinoshita2016summary} or the CHiME~\cite{barker2013pascal,barker2015third} challenges. 

In this paper, we propose a new approach for noise-aware training of deep neural network (DNN) based acoustic models in hybrid ASR systems. We leverage a speech activity detection (SAD) model to identify the speech frames and the silence frames in the utterance. Our ``noise vector'', then, is obtained by concatenating the means of the speech frames and the silence frames, respectively. We conjecture that such a noise estimate is well-informed by the distortions occurring throughout the utterance, including additive and multiplicative noises. The use of non-speech frames to estimate the noise in an utterance has also motivated previous work on noise-robust ASR~\cite{seltzer2013investigation} and speech denoising~\cite{Xu2020ListeningTS}. Furthermore, in our estimation strategy, this additional information is acquired at no extra computational cost, since we make use of a GMM-HMM model that is already trained for generating alignments for training the DNN acoustic model.

In online speech recognition, we are only provided a partial input sequence, and the model is required to output streaming transcriptions. In such a setting, waiting until the whole utterance has been seen in order to compute the means of the speech and silence frames is not feasible, thus requiring online estimates of the noise vector. Towards this objective, we propose to use maximum likelihood (ML) estimates obtained from a probabilistic model of the means of speech and silence. This makes our noise-adaptive training scheme compatible with streaming ASR systems.

\section{Related Work}
\label{sec:related}

In the past, several methods have been proposed to make ASR systems robust to environmental noise~\cite{li2014overview}, either through feature normalization (like spectral subtraction~\cite{boll1979suppression} and Weiner filtering~\cite{loizou2013speech}) or through model adaptation, such as MLLR-like schemes~\cite{leggetter1995maximum}. Recently, with the ubiquitous adoption of neural networks in hybrid as well as end-to-end ASR, data-driven supervised approaches have emerged as potential alternatives to traditional unsupervised methods~\cite{zhang2018deep}. These include front-end enhancement techniques like masking~\cite{srinivasan2006binary} or back-end methods like noise-aware and multi-condition training~\cite{seltzer2013investigation}. 

For robust streaming ASR, several online front-end processing methods have been proposed, such as time-frequency masks and steering vectors~\cite{higuchi2016robust} or online-enabled Generalized Eigenvalue (GEV) beamformers~\cite{kitza2016robust}. In this work, we do not perform any front-end processing of the acoustic signals; instead, we obtain an informed estimate of the noise from the observed chunks of speech and non-speech frames, and use this estimate to adapt the acoustic model.

Several researchers have proposed the use of utterance-level embeddings for noise/environment adaptation of acoustic models. Saon et al.~\cite{saon2013speaker} used i-vectors, proposed originally for speaker verification~\cite{Dehak2011FrontEndFA}. An ``online'' version of these i-vectors, estimated every 10 frames, is supported in the Kaldi~\cite{povey2011kaldi} ASR toolkit. Seltzer et al.~\cite{seltzer2013investigation} used noise-aware training similar to our approach, but their noise vectors were estimated by averaging the first and last 10 frames in the utterance, whereas we estimate noise by leveraging a SAD module. In \cite{Kim2016EnvironmentalNE}, the authors trained a network to predict the environment label, and used the embedding from the bottleneck-layer to inform the acoustic model. Similarly, \cite{Feng2017AnEF} trained a linear discriminant analysis (LDA) model using environment labels to project i-vectors into a space that more accurately represents noise variability in the utterance, calling them e-vectors. They also used a bottleneck DNN similar to~\cite{Kim2016EnvironmentalNE} for estimating e-vectors, with the exception that they used i-vectors as input, instead of the acoustic features. We will look at some of these approaches in more detail in Section~\ref{sec:baselines}.

Since we use the means of acoustic features, comparisons may also be drawn to the popular cepstral mean normalization (CMN)~\cite{liu-etal-1993-efficient} approach. However, while CMN just subtracts the cepstral mean, our method allows the DNN to learn non-linear transforms of the noise vector.

\section{Noise Vectors}
\label{sec:noise}

We make the acoustic model noise-aware by appending additional information with the input feature sequence in the form of ``noise vectors.'' Let $\xvec_i=(\xvec_{i1},\ldots,\xvec_{it},\ldots,\xvec_{iT_i})$, $\xvec_{it}\in \mathbb{R}^d$, be the input feature frames corresponding to the utterance $i$. Consider an oracle SAD function, $f:\mathbb{R}^d \rightarrow \{\text{speech},\text{sil}\}, f(\xvec_{it}) = s_{it}$, which classifies a frame as speech or silence. Let  $\svec_i=(s_{i1},\ldots,s_{it},\ldots,s_{iT_i})$ be the corresponding SAD predictions for the input sequence. We compute the means of speech and silence frames, $\musi$ and $\muni$, as
\begin{equation}
\label{eqn:noise_vec}
\musi = \frac{\sum_{t \in \text{speech}} \xvec_{it}} {\sum_{t \in \text{speech}} 1},~~\text{and}~~
\muni =\frac{\sum_{t \in \text{sil}} \xvec_{it}} {\sum_{t \in \text{sil}} 1}.
\end{equation}

\begin{figure}[t]
\centering
    \includegraphics[width=\linewidth]{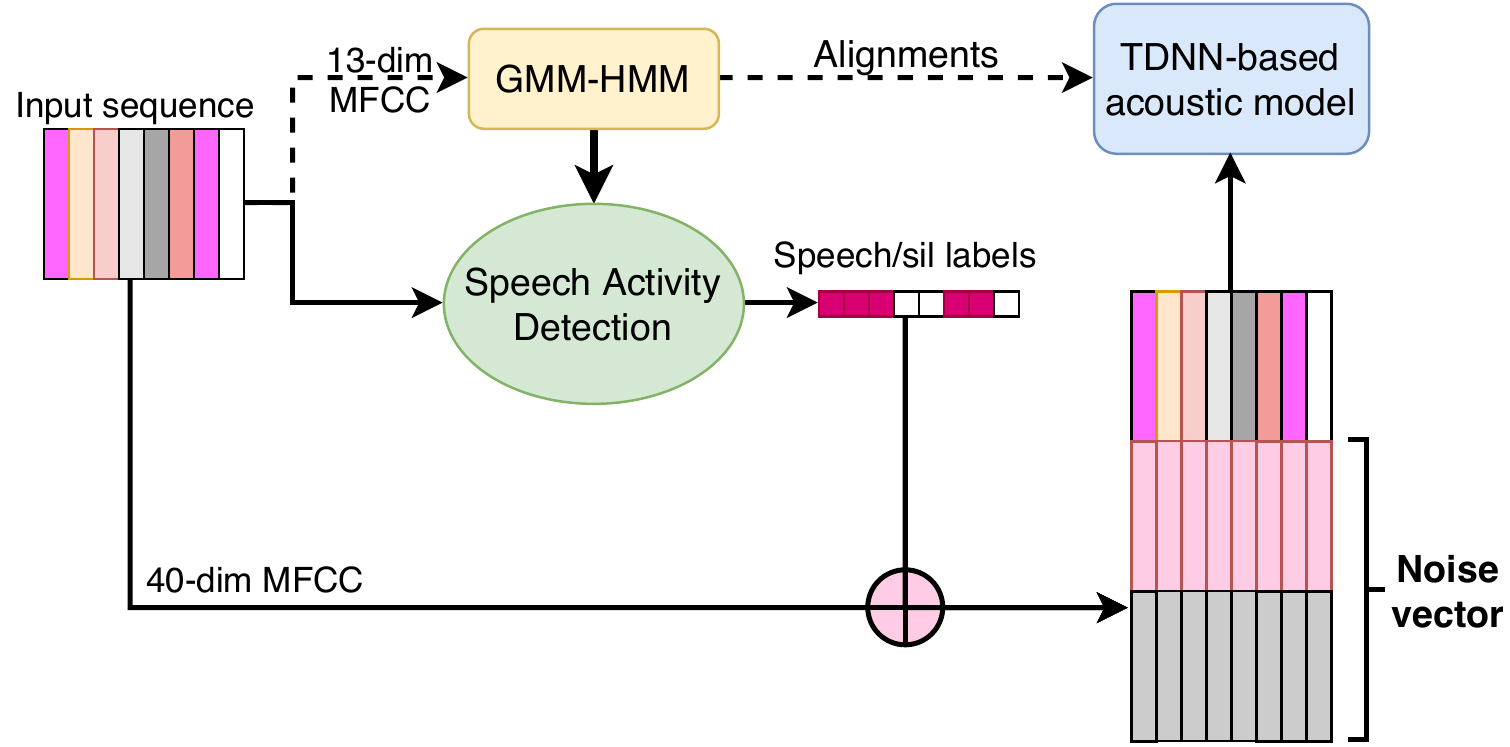}
    \caption{Noise-aware training/inference pipeline with a TDNN-based acoustic model. Dotted lines denote paths that are only required during training.}
    \label{fig:overview}
\end{figure}

In our hybrid ASR system, the acoustic model is trained using a sequence discriminative criterion~\cite{vesely2013sequence} such as lattice-free MMI~\cite{povey2016purely}, on alignments generated using a GMM-HMM model with the training transcriptions. Since we know which phones map to silence phones, we use the same alignments to obtain the SAD label sequence $\svec_i$. An overview of our noise-aware training scheme is shown in Fig.~\ref{fig:overview}. We append the noise vectors to the input features through a linear transformation (called a ``control layer'' in literature~\cite{Rownicka2019EmbeddingsFD})\footnote{This is conceptually similar to an LDA feature transformation~\cite{Somervuo2003FeatureTA}.}.

At inference time, we run a first-pass decoding with the same GMM-HMM model to obtain the decoding lattice, which is then used to obtain the SAD labels for estimating the noise vectors. Since we only need approximate representations of speech and noise frames, the classification need not be perfect, and so this strategy suffices. Next, we describe an  extension of this estimation technique to the online decoding scenario using maximum likelihood estimate.

\subsubsection*{Maximum likelihood (ML) estimate}
The ML estimate of the noise vectors is similar to equation~\eqref{eqn:noise_vec}, with $\svec_i$ being the partial sequence observed until time $t$. If no speech or silence frames are observed, the corresponding components are set to 0. 
In Section~\ref{sec:mle}, we will show that this estimate of the noise vectors quickly converges to the offline estimate in most settings.\footnote{In addition to this maximum likelihood estimate of the noise vectors, we can also derive a \emph{maximum a posteriori} (MAP) estimate which ties together the characteristics of both the speech and silence classes. The derivation is given in Appendix~\ref{appendix:map}. In our preliminary experiments, however, this model did not show any improvements over the simple ML estimate.}

\section{Experimental setup}

\subsection{Datasets and model}

We performed experiments on two different datasets: AMI and Aurora-4. AMI~\cite{carletta2005ami} consists of 100 hours of recorded meetings, and speech from close-talk, single distant microphone (SDM), and array microphones is provided. We used the SDM recordings for our experiments, which is ideal for single-channel far-field evaluation. Aurora-4~\cite{yeung2004improved} is a 14-hour close-talk speech corpus based on Wall Street Journal (WSJ0) data with artificially added noise. We evaluated on the \texttt{eval92} and \texttt{0166} subsets. All our experiments were conducted using the Kaldi speech recognition toolkit~\cite{povey2011kaldi} using neural networks trained with lattice-free MMI~\cite{povey2016purely}. The acoustic model for Aurora-4 was based on factored TDNN~\cite{povey2018semi}, while the one for AMI was TDNN-based. The hyperparameters for each setup are shown in Table~\ref{tab:am_details}. For the Aurora-4 setup, we used a pruned 3-gram language model (LM) trained on WSJ transcriptions using the IRSTLM toolkit~\cite{federico2008irstlm}. For AMI, we trained a pruned 3-gram LM with KN-smoothing on a combination of AMI and Fisher~\cite{cieri2004fisher} data using the SRILM toolkit~\cite{stolcke2002srilm}.


\begin{table}[t]
\centering
\caption{Hyperparameters for the acoustic model training for our experimental setups.}
\label{tab:am_details}
\begin{adjustbox}{max width=\linewidth}
\begin{tabular}{@{}llcccc@{}}
\toprule
\textbf{Setup} & \textbf{Model} & \textbf{\# layers} & \textbf{Dim} & \textbf{Bottleneck} & \textbf{\# epochs} \\
\midrule
AMI & TDNN & 9 & 450 & - & 9 \\
Aurora-4 & TDNN-F & 12 & 1024 & 128 & 10 \\
\bottomrule
\end{tabular}
\end{adjustbox}
\end{table}

\subsection{Baselines}
\label{sec:baselines}

As described earlier in Section~\ref{sec:related}, several embedding-based methods have been proposed for noise-aware training of acoustic models. Here, we perform comparisons with 4 of these methods, described below.

\begin{enumerate}[wide, labelwidth=!, labelindent=0pt]
    \item \textbf{i-vector}: i-vectors were first introduced in \cite{Dehak2011FrontEndFA} for speaker verification, and later used for speaker adaptation of acoustic models in \cite{saon2013speaker}. They are generative embeddings which model both speaker and channel variability in utterances. We used two variants in our experiments: \textit{offline}, which were estimated over the whole utterance, and \textit{online}, which were estimated every 10 frames. For both these variants, we extracted 100-dim i-vectors using extractors trained on the training data.
    
    \item \textbf{NAT-vector}: Seltzer et al.~\cite{seltzer2013investigation} proposed ``noise vectors'' for noise-aware training, by averaging the first and last 10 frames of the utterance. We refer to these vectors as NAT-vectors, to avoid confusion with our proposed noise vectors. Since we use 40-dim MFCC features for training our models, these vectors are also 40-dimensional.
    
    \item \textbf{e-vector}: Since i-vectors capture both speaker and channel variability, \cite{Feng2017AnEF} proposed a technique for extracting the ``environment'' variability by transforming the i-vectors using environment labels, either using linear discriminant analysis (LDA), or bottleneck neural networks (NNs), and called them e-vectors. Here, we use the LDA variant of e-vectors for our comparison. For Aurora-4, we trained the e-vectors on 14 noise label types from the official development set. Since AMI does not have annotated environment labels, we trained the LDA on CHiME-4 data~\cite{barker2015third}, which consists of 4 different noise types. For both the datasets, we transformed the 100-dim i-vectors to 50-dim e-vectors.
    
    \item \textbf{Bottleneck NN}: Since the above baselines estimate environment embeddings at the utterance level, Kim et al.~\cite{Kim2016EnvironmentalNE} argued that they cannot effectively capture non-stationary noise. Instead, they trained bottleneck NNs to classify noise types with frame-level acoustic features provided as input. Frame-level noise embeddings were then extracted from the bottleneck layer and used for noise-aware training. For our experiments, we trained a 5-layer feed-forward neural network with 1024 hidden units (similar to \cite{Kim2016EnvironmentalNE}), and a 80-dim bottleneck was placed at the 4th layer. Noise labels used as training targets for the bottleneck NNs were obtained similar to the e-vector LDA training.
\end{enumerate}

In addition to these embedding methods from literature, we report results using (i) CMN, and (ii) embeddings obtained by averaging the entire utterance, which we call \texttt{utt-mean}. This ablation is performed to demonstrate the importance of obtaining separate means for speech and silence frames. Our implementation for these baselines and our proposed method will be made publicly available at: {\footnotesize\href{https://github.com/desh2608/kaldi-noise-vectors}{\texttt{https://github.com/desh2608/kaldi-noise-vectors}}}.

\section{Results and Discussion}

\subsection{Comparison with baselines}

\begin{table}[t]
\centering
\caption{WER results for our proposed noise vectors (offline and MLE) compared with other embedding-based techniques from literature. Methods marked with $^\dagger$ require environment labels for training embedding extractor.}
\label{tab:results}
\begin{adjustbox}{width=\linewidth} 
\begin{tabular}{@{}lcccccc@{}}
\toprule
\multirow{2}{*}{\textbf{Method}} & \phantom{a} & \multicolumn{2}{c}{\textbf{AMI}} & \phantom{a} & \multicolumn{2}{c}{\textbf{Aurora-4}} \\ 
 \cmidrule(l{2pt}r{2pt}){3-4} \cmidrule(l{2pt}r{2pt}){6-7}
  && \textbf{Dev} & \textbf{Eval} && \textbf{Eval-92} & \textbf{0166} \\ 
\midrule
Base model &&  38.8 & 42.6 && 7.94 & 8.12 \\
CMN && 38.5 & 42.8 && 7.79 & 7.94 \\
utt-mean && 36.8 & 40.8 && 7.77 & 7.78 \\
\midrule
i-vector (offline)~\cite{saon2013speaker} &&  35.2  & 39.4 && 8.24 & 8.53 \\
i-vector (online) &&  35.6 & 40.0 && 8.51 & 8.78 \\
NAT-vector~\cite{seltzer2013investigation} && 38.1 & 42.0 && 8.17 & 8.38  \\
e-vector (LDA)$^{\dagger}$~\cite{Feng2017AnEF} && 37.6 & 41.5 && 7.63 & 7.73 \\
Bottleneck NN$^{\dagger}$~\cite{Kim2016EnvironmentalNE} && 38.7 & 42.4 && 7.77 & 7.98 \\
\midrule
Noise vector (offline) &&  \textbf{34.9} & \textbf{38.8} && \textbf{7.37} & \textbf{7.61} \\
Noise vector (MLE) &&  35.4 & 39.2 && 7.72 & 7.93 \\ 
\bottomrule
\end{tabular}
\end{adjustbox}
\end{table}

\begin{table}[t]
\centering
\caption{Insertion (ins), deletion (del), and substitution (sub) error comparison for noise vectors compared with the base model, for AMI (dev) and Aurora-4 (eval92).}
\label{tab:breakdown}
\begin{adjustbox}{width=0.9\linewidth} 
\begin{tabular}{@{}llcccc@{}}
\toprule
\textbf{Dataset} & \textbf{System} & \textbf{Ins} & \textbf{Del} & \textbf{Sub} & \textbf{WER} \\
\midrule
\multirow{2}{*}{AMI (dev)} & TDNN & 4.7 & 12.6 & 21.5 & 38.8 \\
& + noise vector &  3.2 & 11.5 & 20.3 & 34.9 \\
\midrule
\multirow{2}{*}{\shortstack[l]{Aurora-4\\(eval92)}} & TDNN-F & 0.50 & 2.57 & 4.87 & 7.94 \\
& + noise vector & 0.58 & 2.12 & 4.67 & 7.37 \\
\bottomrule
\end{tabular}
\end{adjustbox}
\end{table}


\begin{figure*}[t]
\begin{subfigure}{\linewidth}
\centering
\cutimage{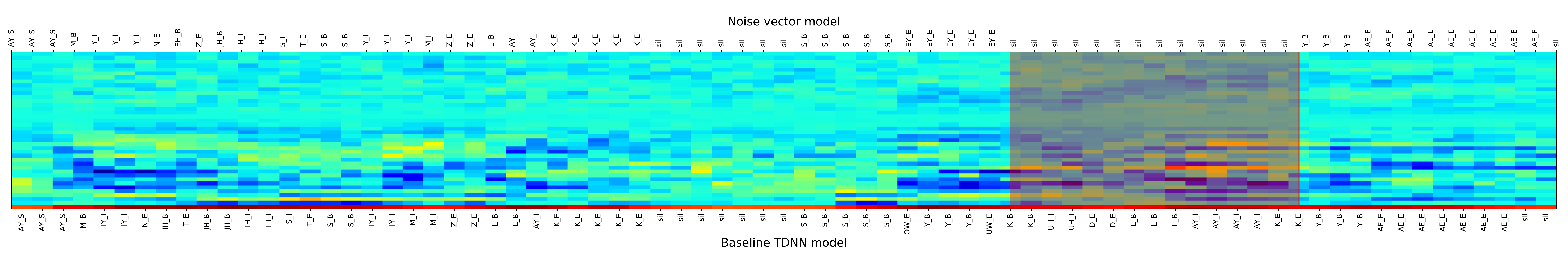}{0.65}{0.65}
\end{subfigure}\hfill
\vspace{-1em}
\begin{subfigure}{\linewidth}
\centering
\cutimage{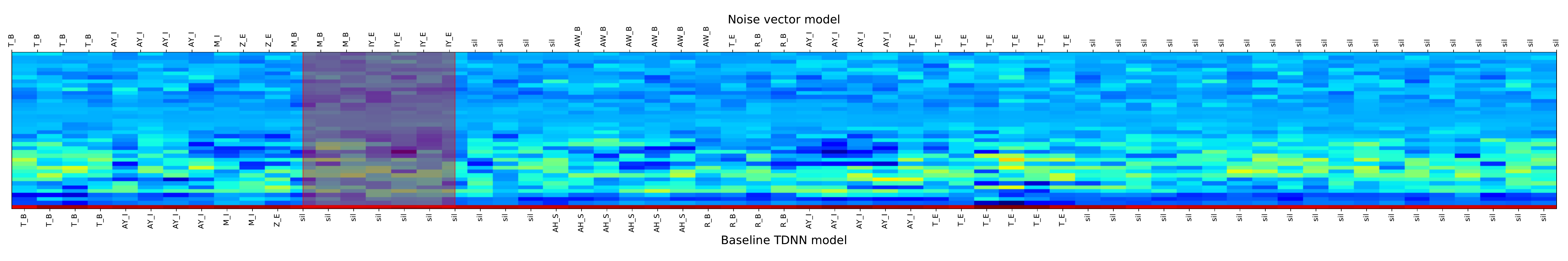}{0.65}{0.65}
\end{subfigure}\hfill
\vspace{-1em}
\caption{Frame-level phone alignments obtained from the Baseline TDNN and noise vector models for the sentences: {\normalfont \texttt{I MEAN IT JUST SEEMS LIKE YEAH}} (top) and  {\normalfont \texttt{TILL THE MEETING OH RIGHT}} (bottom) in the AMI Dev set. The shaded region shows where the baseline model makes insertion and deletion errors, respectively. We have only shown part of the spectograms for each utterance.}
\label{fig:analysis}
\end{figure*}

Table~\ref{tab:results} shows the WER results for the proposed method and the baselines on AMI and Aurora-4. It can be seen from the table that adding offline noise vectors to the input features provided WER improvements of 10\% and 7.1\% on AMI and Aurora-4, respectively. Furthermore, these vectors consistently outperformed the \texttt{utt-mean} baseline, indicating that the speech/silence classification is important for the model. MLE-based vectors are also effective, although they perform slightly worse than the utterance-level vectors, as expected. We conjecture that this may be because of incorrect estimates of speech and silence in the beginning of the utterance.

Among the baseline embedding-based systems, i-vectors (both offline and online) showed significant gains for AMI, but degraded results on Aurora-4. This may be because Aurora-4 utterances contain long silences, which can corrupt i-vector estimation. NAT vectors also followed the same trend, indicating that noise estimation from only the first and last few frames leads to inaccuracies. On the other hand, e-vectors and bottleneck NNs improved WERs over the base models on both datasets. Our proposed method provided the best performance across the board, demonstrating its robustness. In the next section, we will analyze the performance of our noise vectors further using some ablation experiments.

\begin{table}[t]
\centering
\caption{Effect of speech and non-speech frames in the noise vector.}
\label{tab:speech_sil}
\begin{adjustbox}{width=0.85\linewidth} 
\begin{tabular}{@{}cccc@{}}
\toprule
\textbf{Speech mean} & \textbf{Non-speech mean} & \textbf{AMI Dev} & \textbf{AMI Eval} \\
\midrule
\xmark & \xmark &  38.8 & 42.6 \\
\xmark & \cmark & 38.1 & 41.9 \\
\cmark & \xmark & 36.7 & 40.8 \\
\cmark & \cmark &  34.9 & 38.8 \\
\bottomrule
\end{tabular}
\end{adjustbox}
\end{table}

\subsection{Where does the improvement come from?}

\textbf{WER breakdown by error type.} Table~\ref{tab:breakdown} shows the insertion, deletion, and substitution error rates for the proposed method. On the AMI dataset, which contains late reverberations and interfering speakers, our noise vectors reduce insertion errors substantially, suggesting that utterance-level characteristics help in speaker selection. In Fig.~\ref{fig:analysis} (top), we see that the baseline model inserted extraneous phonemes between \texttt{LIKE} and \texttt{YEAH}, as a result of cross-talk from the interfering speaker, but the noise-aware model correctly identified the frames as silence. Since Aurora-4 does not contain far-field or overlapping speech, noise vectors do not improve insertion errors on that setting. On both datasets, noise vectors consistently reduced deletion and insertion errors. Qualitative analysis (such as that in Fig.~\ref{fig:analysis} (bottom)) showed that this was a result of better recognition on noisy frames, where the base model did not predict any phones, or predicted incorrect phones.

\textbf{Effect of speech and non-speech frames.} In Table~\ref{tab:results}, comparison with the \texttt{utt-mean} baseline showed that averaging speech and non-speech frames separately was important for the effectiveness of our noise vectors. To measure the individual contributions of these components, we conducted experiments on the AMI dataset by selecting each of the components alternatively (Table~\ref{tab:speech_sil}). We found that speech frames are more important than non-speech frames, but their combination results in the best performance. The relatively smaller WER improvement using non-speech frames is also consistent with the NAT vector results in Table~\ref{tab:results}.

\begin{figure}[t]
\centering
    \includegraphics[width=\linewidth]{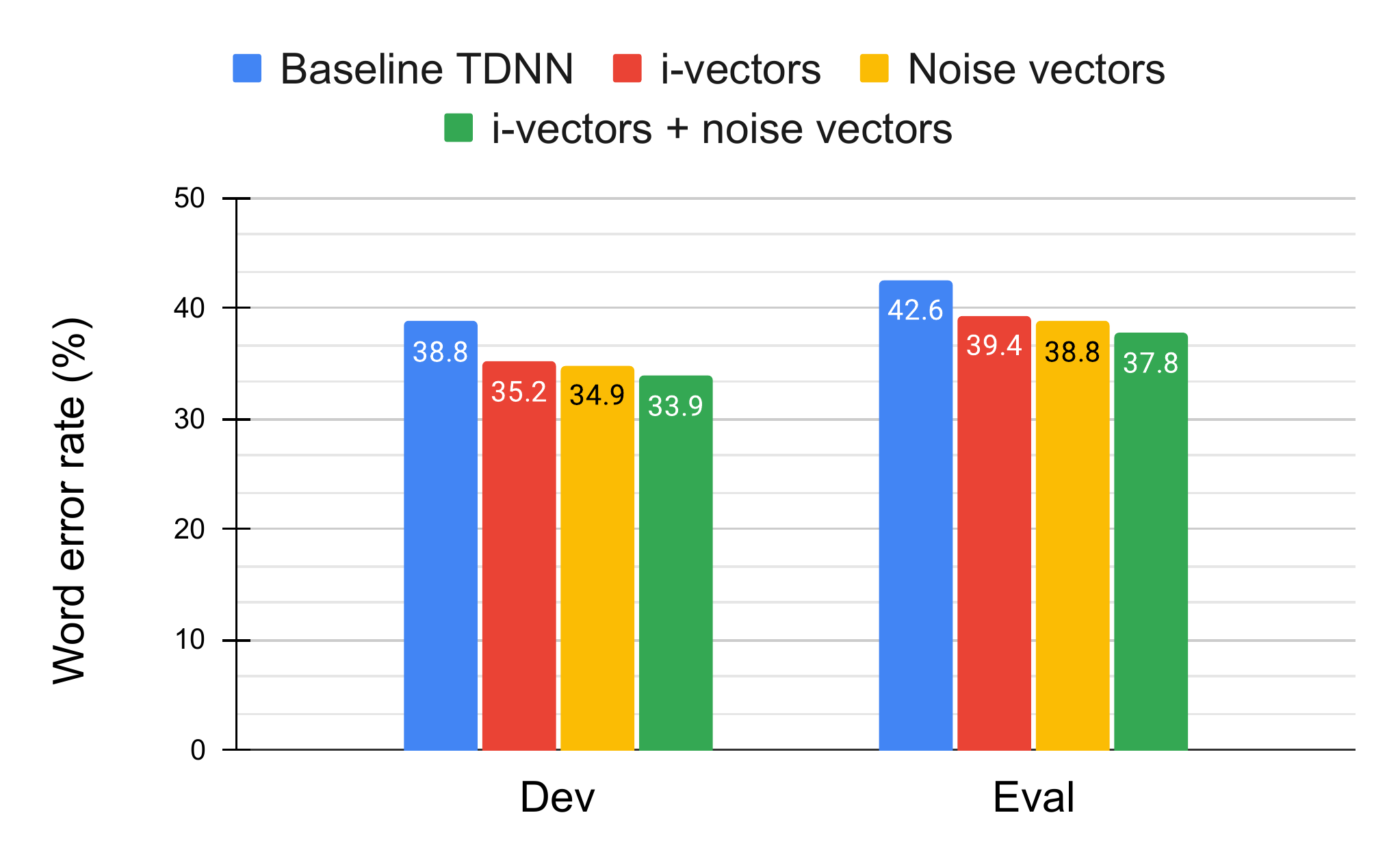}
    \vspace{-1em}
    \caption{Effect of combining i-vectors and noise vectors on the AMI dataset. Noise vectors outperform i-vectors, but they are complementary.}
    \label{fig:ivec_vs_nvec}
\end{figure}

\textbf{Combination with i-vectors.} In Fig.~\ref{fig:ivec_vs_nvec}, we show the effect of combining i-vectors and noise vectors on the AMI dataset. We found that the WER gains using these embedding methods are complementary, and their combination achieved a relative improvement of approx. 12.8\% compared with the baseline TDNN model. This is likely because i-vectors and noise vectors capture speaker and environment characteristics in the utterance, respectively.

\textbf{Combination with multi-condition training.} We also compared our method with multi-condition (MC) training on Aurora-4, where we applied speed and volume perturbation, and added simulated reverberation and point-source noises to the training data, resulting in 6x the original training set. MC training outperformed noise vectors; however, their combination achieved the best performance, reducing the WER by approx. 21.6\% relative over the baseline TDNN-F model. 

\begin{figure}[t]
\centering
    \includegraphics[width=\linewidth]{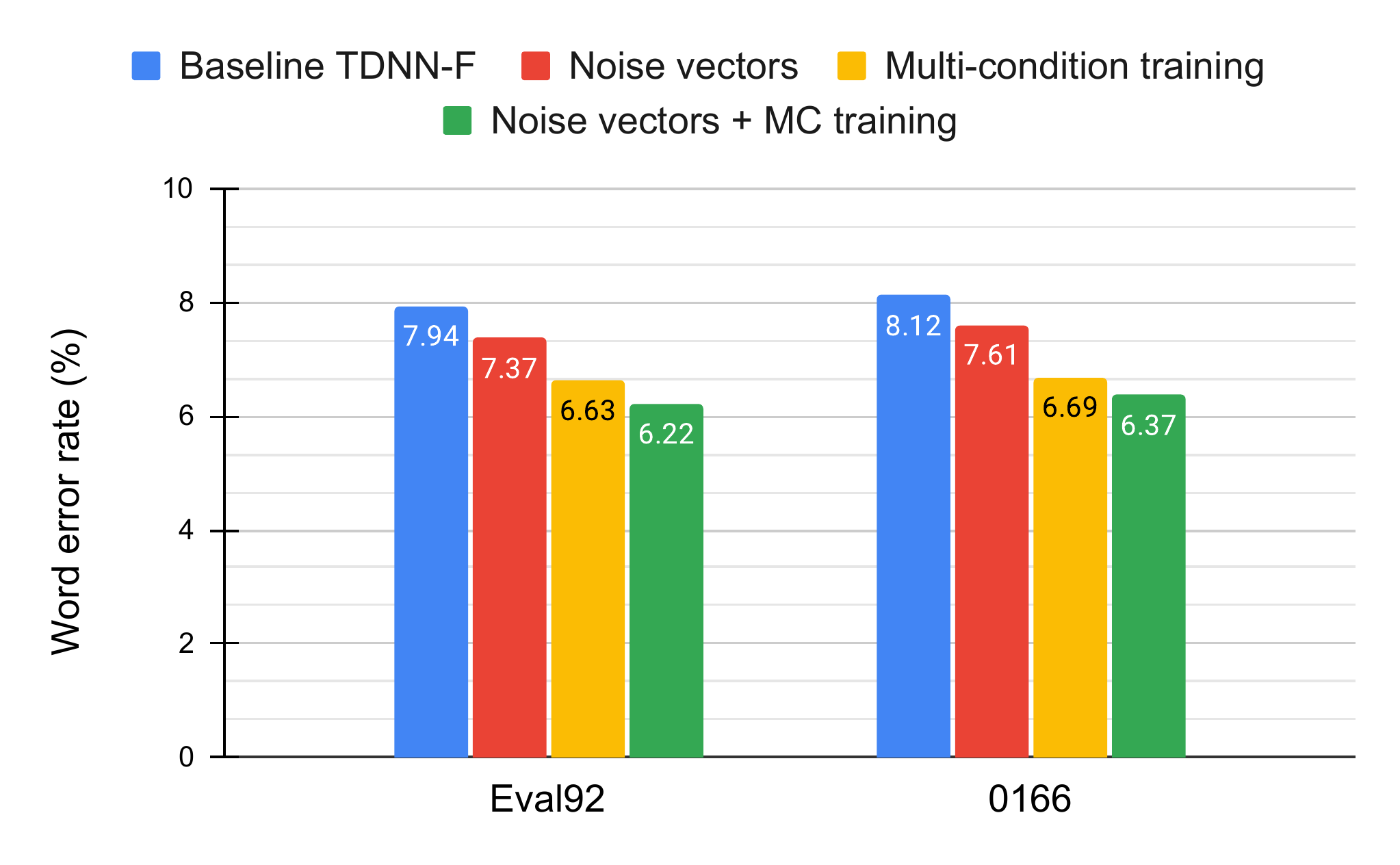}
    \vspace{-1em}
    \caption{Effect of combining noise vectors with multi-condition (MC) training on the Aurora-4 dataset. MC training outperforms noise vectors, but they are complementary.}
    \label{fig:nvec_vs_mc}
\end{figure}

\subsection{Offline vs. MLE noise vectors}
\label{sec:mle}

\begin{figure}[t]
\centering
\includegraphics[width=\linewidth]{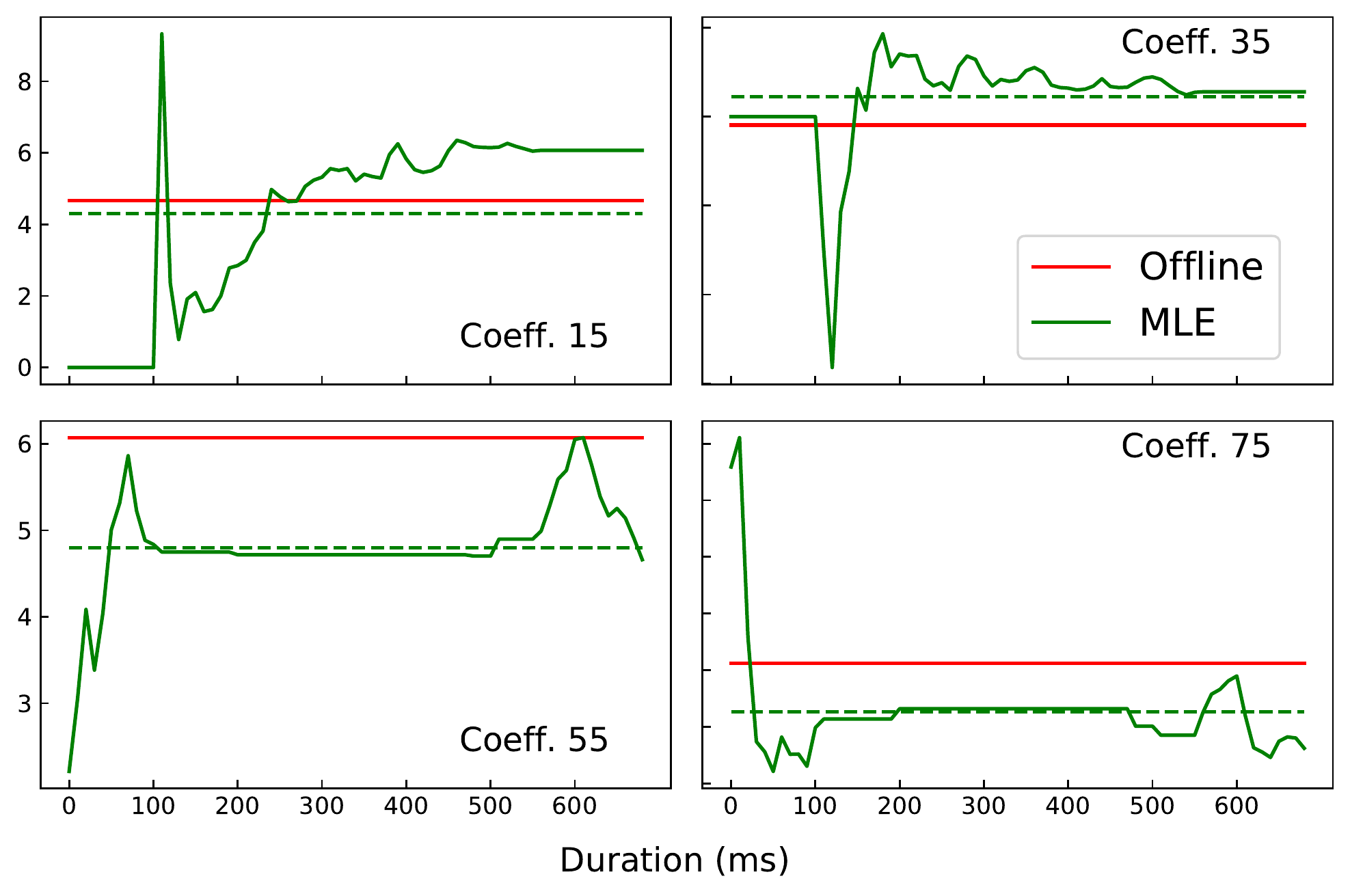}
\caption{Online noise estimates using MLE and MAP for a randomly selected utterance in the Aurora-4 {\normalfont\texttt{test\_0166}} set. The top and bottom row coefficients correspond to means of speech and silence frames, respectively. For each coefficient, the solid red and green lines are the ``offline'' and MLE estimates, respectively. The dashed green line represents the mean of the MLE noise vector over the whole utterance.}
\label{fig:online_analysis}
\end{figure}

Fig.~\ref{fig:online_analysis} shows 4 (out of 80) coefficients from the offline and MLE noise vectors for an utterance selected at random from the Aurora-4 \texttt{test\_0166} set. We see that the online estimates for speech coefficients (15 and 35) quickly approached the utterance-level values. For the noise coefficients (55 and 75), the utterance-level averages for the online estimates were higher because of fewer silence frames in the middle of the utterance, but the overall estimate over the whole utterance was close to the offline vector. Similar observations were made across both datasets.

\section{Conclusions}

We proposed a simple method that leverages a SAD to perform noise-aware training of acoustic models in hybrid ASR systems by computing the means of speech and silence frames. Our method, which we call ``noise vectors,'' provided significant improvements in WER in diverse conditions such as noise and far-field reverberant speech, as demonstrated through experiments conducted on Aurora-4 and AMI. On both conditions, our method outperformed strong baselines that use i-vectors, e-vectors trained with LDA, or bottleneck neural networks. 
Through error analysis, we hypothesized that the improvements are a results of better performance of our model on frames containing high background noise or interference in the form of overlapping speech. Finally, we extended our method to the case of streaming ASR by showing that similar WER gains can be obtained using maximum likelihood estimates of the noise vectors. 

\bibliographystyle{IEEEtran}
{\small
\bibliography{mybib}

\begin{thebibliography}{10}
\providecommand{\url}[1]{#1}
\csname url@samestyle\endcsname
\providecommand{\newblock}{\relax}
\providecommand{\bibinfo}[2]{#2}
\providecommand{\BIBentrySTDinterwordspacing}{\spaceskip=0pt\relax}
\providecommand{\BIBentryALTinterwordstretchfactor}{4}
\providecommand{\BIBentryALTinterwordspacing}{\spaceskip=\fontdimen2\font plus
\BIBentryALTinterwordstretchfactor\fontdimen3\font minus
  \fontdimen4\font\relax}
\providecommand{\BIBforeignlanguage}[2]{{%
\expandafter\ifx\csname l@#1\endcsname\relax
\typeout{** WARNING: IEEEtran.bst: No hyphenation pattern has been}%
\typeout{** loaded for the language `#1'. Using the pattern for}%
\typeout{** the default language instead.}%
\else
\language=\csname l@#1\endcsname
\fi
#2}}
\providecommand{\BIBdecl}{\relax}
\BIBdecl

\bibitem{kinoshita2016summary}
K.~Kinoshita, M.~Delcroix, S.~Gannot, E.~A. Habets, R.~Haeb-Umbach,
  W.~Kellermann, V.~Leutnant, R.~Maas, T.~Nakatani, B.~Raj \emph{et~al.}, ``A
  summary of the {REVERB} challenge: state-of-the-art and remaining challenges
  in reverberant speech processing research,'' \emph{EURASIP Journal on
  Advances in Signal Processing}, vol. 2016, no.~1, p.~7, 2016.

\bibitem{barker2013pascal}
J.~Barker, E.~Vincent, N.~Ma, H.~Christensen, and P.~Green, ``The {PASCAL}
  {CHiME} speech separation and recognition challenge,'' \emph{Computer Speech
  \& Language}, vol.~27, no.~3, pp. 621--633, 2013.

\bibitem{barker2015third}
J.~Barker, R.~Marxer, E.~Vincent, and S.~Watanabe, ``The third ‘{CHiME}’
  speech separation and recognition challenge: Dataset, task and baselines,''
  in \emph{IEEE ASRU}.\hskip 1em plus 0.5em minus 0.4em\relax IEEE, 2015, pp.
  504--511.

\bibitem{seltzer2013investigation}
M.~L. Seltzer, D.~Yu, and Y.~Wang, ``An investigation of deep neural networks
  for noise robust speech recognition,'' in \emph{IEEE ICASSP}.\hskip 1em plus
  0.5em minus 0.4em\relax IEEE, 2013, pp. 7398--7402.

\bibitem{Xu2020ListeningTS}
R.-L. Xu, R.~Wu, Y.~Ishiwaka, C.~Vondrick, and C.~Zheng, ``Listening to sounds
  of silence for speech denoising,'' \emph{ArXiv}, vol. abs/2010.12013, 2020.

\bibitem{li2014overview}
J.~Li, L.~Deng, Y.~Gong, and R.~Haeb-Umbach, ``An overview of noise-robust
  automatic speech recognition,'' \emph{IEEE/ACM Transactions on Audio, Speech,
  and Language Processing}, vol.~22, no.~4, pp. 745--777, 2014.

\bibitem{boll1979suppression}
S.~Boll, ``Suppression of acoustic noise in speech using spectral
  subtraction,'' \emph{IEEE Transactions on acoustics, speech, and signal
  processing}, vol.~27, no.~2, pp. 113--120, 1979.

\bibitem{loizou2013speech}
P.~C. Loizou, \emph{Speech enhancement: theory and practice}.\hskip 1em plus
  0.5em minus 0.4em\relax CRC press, 2013.

\bibitem{leggetter1995maximum}
C.~J. Leggetter and P.~C. Woodland, ``Maximum likelihood linear regression for
  speaker adaptation of continuous density hidden markov models,''
  \emph{Computer speech \& language}, vol.~9, no.~2, pp. 171--185, 1995.

\bibitem{zhang2018deep}
Z.~Zhang, J.~Geiger, J.~Pohjalainen, A.~E.-D. Mousa, W.~Jin, and B.~Schuller,
  ``Deep learning for environmentally robust speech recognition: An overview of
  recent developments,'' \emph{ACM Transactions on Intelligent Systems and
  Technology (TIST)}, vol.~9, no.~5, pp. 1--28, 2018.

\bibitem{srinivasan2006binary}
S.~Srinivasan, N.~Roman, and D.~Wang, ``Binary and ratio time-frequency masks
  for robust speech recognition,'' \emph{Speech Communication}, vol.~48,
  no.~11, pp. 1486--1501, 2006.

\bibitem{higuchi2016robust}
T.~Higuchi, N.~Ito, T.~Yoshioka, and T.~Nakatani, ``Robust {MVDR} beamforming
  using time-frequency masks for online/offline asr in noise,'' in \emph{IEEE
  ICASSP}.\hskip 1em plus 0.5em minus 0.4em\relax IEEE, 2016, pp. 5210--5214.

\bibitem{kitza2016robust}
M.~Kitza, A.~Zeyer, R.~Schl{\"u}ter, J.~Heymann, and R.~Haeb-Umbach, ``Robust
  online multi-channel speech recognition,'' in \emph{Speech Communication; 12.
  ITG Symposium}.\hskip 1em plus 0.5em minus 0.4em\relax VDE, 2016.

\bibitem{saon2013speaker}
G.~Saon, H.~Soltau, D.~Nahamoo, and M.~Picheny, ``Speaker adaptation of neural
  network acoustic models using i-vectors,'' in \emph{IEEE ASRU}.\hskip 1em
  plus 0.5em minus 0.4em\relax IEEE, 2013, pp. 55--59.

\bibitem{Dehak2011FrontEndFA}
N.~Dehak, P.~Kenny, R.~Dehak, P.~Dumouchel, and P.~Ouellet, ``Front-end factor
  analysis for speaker verification,'' \emph{IEEE Transactions on Audio,
  Speech, and Language Processing}, vol.~19, pp. 788--798, 2011.

\bibitem{povey2011kaldi}
D.~Povey, A.~Ghoshal, G.~Boulianne, L.~Burget, O.~Glembek, N.~Goel,
  M.~Hannemann, P.~Motlicek, Y.~Qian, P.~Schwarz \emph{et~al.}, ``The {Kaldi}
  speech recognition toolkit,'' in \emph{IEEE ASRU}, no. CONF.\hskip 1em plus
  0.5em minus 0.4em\relax IEEE Signal Processing Society, 2011.

\bibitem{Kim2016EnvironmentalNE}
S.~Kim, B.~Raj, and I.~Lane, ``Environmental noise embeddings for robust speech
  recognition,'' \emph{ArXiv}, vol. abs/1601.02553, 2016.

\bibitem{Feng2017AnEF}
X.~Feng, B.~Richardson, S.~Amman, and J.~R. Glass, ``An environmental feature
  representation for robust speech recognition and for environment
  identification,'' in \emph{INTERSPEECH}, 2017.

\bibitem{liu-etal-1993-efficient}
\BIBentryALTinterwordspacing
F.-H. Liu, R.~M. Stern, X.~Huang, and A.~Acero, ``Efficient cepstral
  normalization for robust speech recognition,'' in \emph{{H}uman {L}anguage
  {T}echnology: Proceedings of a Workshop Held at Plainsboro, New Jersey, March
  21-24, 1993}, 1993. [Online]. Available:
  \url{https://www.aclweb.org/anthology/H93-1014}
\BIBentrySTDinterwordspacing

\bibitem{vesely2013sequence}
K.~Vesel{\`y}, A.~Ghoshal, L.~Burget, and D.~Povey, ``Sequence-discriminative
  training of deep neural networks.'' in \emph{INTERSPEECH}, vol. 2013, 2013,
  pp. 2345--2349.

\bibitem{povey2016purely}
D.~Povey, V.~Peddinti, D.~Galvez, P.~Ghahremani, V.~Manohar, X.~Na, Y.~Wang,
  and S.~Khudanpur, ``Purely sequence-trained neural networks for {ASR} based
  on lattice-free {MMI}.'' in \emph{INTERSPEECH}, 2016, pp. 2751--2755.

\bibitem{Rownicka2019EmbeddingsFD}
J.~Rownicka, P.~Bell, and S.~Renals, ``Embeddings for {DNN} speaker adaptive
  training,'' \emph{IEEE ASRU}, pp. 479--486, 2019.

\bibitem{Somervuo2003FeatureTA}
P.~Somervuo, B.~Y. Chen, and Q.~Zhu, ``Feature transformations and combinations
  for improving {ASR} performance,'' in \emph{INTERSPEECH}, 2003.

\bibitem{carletta2005ami}
J.~Carletta, S.~Ashby, S.~Bourban, M.~Flynn, M.~Guillemot, T.~Hain, J.~Kadlec,
  V.~Karaiskos, W.~Kraaij, M.~Kronenthal \emph{et~al.}, ``The {AMI} meeting
  corpus: A pre-announcement,'' in \emph{International workshop on machine
  learning for multimodal interaction}.\hskip 1em plus 0.5em minus 0.4em\relax
  Springer, 2005, pp. 28--39.

\bibitem{yeung2004improved}
S.-K.~A. Yeung and M.-H. Siu, ``Improved performance of {Aurora} 4 using {HTK}
  and unsupervised {MLLR} adaptation,'' in \emph{Eighth International
  Conference on Spoken Language Processing}, 2004.

\bibitem{povey2018semi}
D.~Povey, G.~Cheng, Y.~Wang, K.~Li, H.~Xu, M.~Yarmohammadi, and S.~Khudanpur,
  ``Semi-orthogonal low-rank matrix factorization for deep neural networks,''
  in \emph{INTERSPEECH}, 2018, pp. 3743--3747.

\bibitem{federico2008irstlm}
M.~Federico, N.~Bertoldi, and M.~Cettolo, ``{IRSTLM}: an open source toolkit
  for handling large scale language models,'' in \emph{INTERSPEECH}, 2008.

\bibitem{cieri2004fisher}
C.~Cieri, D.~Miller, and K.~Walker, ``The {F}isher corpus: a resource for the
  next generations of speech-to-text,'' in \emph{LREC}, vol.~4, 2004, pp.
  69--71.

\bibitem{stolcke2002srilm}
A.~Stolcke, ``{SRILM}-an extensible language modeling toolkit,'' in
  \emph{Seventh International Conference on Spoken Language Processing}, 2002.

\end{thebibliography}
}
\newpage
\onecolumn
\appendix

\section{\emph{Maximum a posteriori} estimation of noise vectors}
\label{appendix:map}

If we consider a frame-level classification of a noisy utterance into speech and silence frames, the noise characteristics of the utterance may be encapsulated by the silence frames. However, the speech frames contain both speech and background noise, and consequently the mean of speech frames are also informed by the mean of noise frames. Considering this, we propose a model that ties together the statistics of both classes, denoted by $\mathbf{s}$ and $\mathbf{n}$.  Let $\musi$ and $\muni$ (which were defined by \eqref{eqn:noise_vec} for the offline case) be the means of speech and silence which we want to estimate. We write the likelihood of the features and means of utterance $i$ as
\begin{align}
    \label{eq:pxmu_rpi}
    &\Prob{\xvec_i,\musi,\muni|\svec_i,\rsi,\rni,\pi}=
    \prod_{t\in\mathrm{sil}} \Prob{\xvec_{it}|\neg s_{it},\muni,\rni,\pi} \nonumber\\
    &\prod_{t\in\mathrm{sp.}} \Prob{\xvec_{it}|s_{it},\musi,\rsi,\pi} 
    \Prob{\musi|\muni,\pi}\Prob{\muni|\pi},
\end{align}
where
\begin{align}
   \label{eq:px_s0}
    &\Prob{\xvec_{it}|s_{it}=0,\muni,\rni,\pi} = \Gauss{\xvec_{it}}{\muni}{(\rni\Ln)^{-1}} \\
    \label{eq:px_s1}
    &\Prob{\xvec_{it}|s_{it}=1,\musi,\rsi,\pi} = \Gauss{\xvec_{it}}{\musi}{(\rsi\Ls)^{-1}}\\
    \label{eq:pmusi_mui}
    &\Prob{\musi|\muni,\pi} = \Gauss{\musi}{\avec+\Bmat\muni}{\Ls^{-1}}\\
    \label{eq:pmui}
    &\Prob{\muni|\pi}=\Gauss{\muni}{\mun}{\Ln^{-1}}
\end{align}
where we defined the prior parameters $\pi=(\mun,\avec,\Bmat,\Ln,\Ls)$. The scaling factors
$\rsi$ and $\rni$ are deterministic parameters that can be set manually or estimated by maximum likelihood from the training data.

\subsection{Estimating the prior}

Assuming that we have access to a large number of frames corresponding to each class at training time, we just estimate $\musi$ and $\muni$ by maximum likelihood and assume that they are observed. Then, for each utterance we can form a vector 
$ \muvec_i = [ \musi , \muni ]^T.$ We calculate the ML mean and covariance of $\muvec_i$ over the whole training set to get the prior joint probability distribution of speech and silence means
$
    \Prob{\muvec_i|\pi} = \Gauss{\muvec_i}{\muvec}{\Sigmat}
$
with
\begin{align}
    \muvec = \begin{bmatrix}
        \mus \\
        \mun
    \end{bmatrix}
    \qquad
   \Sigmat = \begin{bmatrix}
    \Sss & \Ssn\\
    \Sns & \Snn
    \end{bmatrix} \;.
\end{align}
Following, we calculate the precision, $\Lambmat = \iSigmat$. Then, we can use the equations of the marginal and conditional Gaussians\footnote{C.M.Bishop, \emph{Pattern Recognition and Machine Learning}. Springer, 2006.} to obtain the required priors as
\begin{align}
    \label{eq:pmui2}
    &\Prob{\muni|\pi} = \Gauss{\muni}{\mun}{\Snn} \\
    \label{eq:pmusi_mui2}
    &\Prob{\musi|\muni,\pi} = \Gauss{\musi}{\muvec_{\rms|\rmn}}{\Lss^{-1}} \\
    \label{eq:musgivenn}
    & \muvec_{\rms|\rmn} = \mus - \Lss^{-1} \Lsn (\muni - \mun)
\end{align}
Now, comparing~\eqref{eq:pmusi_mui} to~\eqref{eq:pmui} with~\eqref{eq:pmui2} to~\eqref{eq:musgivenn}, we define
$\Ln = \Snn^{-1}$, 
$\Ls = \Lss$,
$\avec = \mus + \Lss^{-1} \Lsn \mun$, and 
$\Bmat = - \Lss^{-1} \Lsn$.

\subsection{Derivation of MAP point estimate}

We want to estimate $\Prob{\musi,\muni|\xvec_i,\svec_i,\rsi,\rni,\pi}$. We define the sufficient statistics as
\begin{align*}
    &\Nsi = \sum_{t\in \mathrm{speech}} 1 & &\Nni = \sum_{t\in \mathrm{sil}} 1 \\
    &\Fsi = \sum_{t\in \mathrm{speech}} \xvec_{it} &  &\Fni = \sum_{t\in \mathrm{sil}} \xvec_{it} \\
    &\Ssi = \sum_{t\in \mathrm{speech}} \xvec_{it}\xvec_{it}^\T &  &\Sni = \sum_{t\in \mathrm{sil}} \xvec_{it}\xvec_{it}^\T
\end{align*}

Using Bayes rule,
\begin{align}
    &\lnProb{\musi,\muni|\xvec_i,\svec_i,\rsi,\rni,\pi} = 
    \sum_{t\in\mathrm{sil}} \lnProb{\xvec_{it}|s_{it}=0,\muni,\rni,\pi} \nonumber\\
    &+ \sum_{t\in\mathrm{speech}} \lnProb{\xvec_{it}|s_{it}=1,\musi,\rsi,\pi} 
    +\lnProb{\musi|\muni,\pi}+\Prob{\muni|\pi} + \const\\
    = & -\frac{\rni}{2}\sum_{t\in\mathrm{sil}} \mahP{\xvec_{it}}{\muni}{\Ln}
    -\frac{\rsi}{2}\sum_{t\in\mathrm{speech}} \mahP{\xvec_{it}}{\musi}{\Ls} \nonumber\\
    &-\med\mahP{\musi-\avec}{\Bmat\muni}{\Ls}-\med\mahP{\muni}{\mun}{\Ln}+\const\\
    =& -\frac{\rni \Nni}{2}\xAx{\muni}{\Ln} + \muni^\T\Ln(\rni\Fni) 
    -\frac{\rsi \Nsi}{2}\xAx{\musi}{\Ls} + \musi^\T\Ls(\rsi\Fsi) \nonumber\\
    & -\med \xAx{\musi}{\Ls} -\med \xAx{\muni}{\Bmat^\T\Ls\Bmat}
    + \musi^\T\Ls\avec + \musi^T\Ls\Bmat\muni + \muni^\T\Bmat^\T\Ls\avec \nonumber\\
    & -\med \xAx{\muni}{\Ln} + \muni^\T\Ln\mun +\const\\
    =& -\frac{1+\rni \Nni}{2}\xAx{\muni}{\Ln} 
    + \muni^\T\left(\Ln(\mun+\rni\Fni)+\Bmat^\T\Ls\avec\right) \nonumber\\
  & -\frac{1+\rsi \Nsi}{2}\xAx{\musi}{\Ls} + \musi^\T\Ls(\avec+\rsi\Fsi) \nonumber\\
    &  -\med \xAx{\muni}{\Bmat^\T\Ls\Bmat}
     + \musi^\T\Ls\Bmat\muni +\const \\
     =& -\med\xAx{\muni}{\left((1+\rni \Nni)\Ln+\Bmat^\T\Ls\Bmat\right)} 
    + \muni^\T\left(\Ln(\mun+\rni\Fni)+\Bmat^\T\Ls\avec\right) \nonumber\\
  & -\frac{1+\rsi \Nsi}{2}\xAx{\musi}{\Ls} + \musi^\T\Ls(\avec+\rsi\Fsi) \nonumber\\
     & -\med \musi^\T(-\Ls\Bmat)\muni  -\med \muni^\T(-\Bmat^\T\Ls)\musi + \const\\
     \label{eq:logpmu_x_final}
     &=-\med \xAx{\muvec_i}{\Kmat_i} + \muvec_i^T \Qmat_i+\const
\end{align}
where we defined
\begin{align}
    \muvec_i =& \begin{bmatrix}
        \musi \\
        \muni
    \end{bmatrix}\\
    \Kmat_i = & \begin{bmatrix}
    (1+\rsi \Nsi)\Ls & -\Ls\Bmat \\
    -\Bmat^\T\Ls & (1+\rni \Nni)\Ln+\Bmat^\T\Ls\Bmat
    \end{bmatrix} \\
    \Qmat_i = & \begin{bmatrix}
    \Ls(\avec+\rsi\Fsi) \\
    \Ln(\mun+\rni\Fni)+\Bmat^\T\Ls\avec
    \end{bmatrix}\:.
\end{align}

Equation~\eqref{eq:logpmu_x_final} has the form of a Gaussian so
\begin{align}
    \Prob{\musi,\muni|\xvec_i,\svec_i,\rsi,\rni,\pi}=&\Gauss{\muvec_i}{\hat{\muvec}_i}{\Kmat_i^{-1}} \\
  \hat{\muvec}_i=& \Kmat_i^{-1}\Qmat_i
\end{align}

$\hat{\muvec}_i$ is the MAP point estimate of means of speech and silence.

\subsection{Estimation of scaling factors}

We need to maximize the EM objective,
\begin{align}
    Q(\rsi,\rni) = \Expcond{\lnProb{\xvec_i,\musi,\muni|\svec_i,\rsi,\rni,\pi}}{\Prob{\musi,\muni|\xvec_i,\svec_i,\rsi,\rni,\pi}}\:.
\end{align}

For $\rsi$,
\begin{align}
    &\frac{\partial Q(\rsi,\rni)}{\partial \rsi} = 
    \sum_{t\in\mathrm{speech}} \frac{\partial}{\partial \rsi}\Exp{\log \Gauss{\xvec_{it}}{\musi}{(\rsi\Ls)^{-1}} }\\
    =& \sum_{t\in\mathrm{speech}} \frac{\partial}{\partial \rsi} \left[\med \log \left|\rsi \Ls\right| 
    - \frac{\rsi}{2}\trace\left(\Exp{\Ls \scattp{\xvec_{it}-\musi}}\right)\right]\\
    =& \frac{\partial}{\partial \rsi} \left[\frac{d\Nsi}{2}\log \rsi 
    - \frac{\rsi}{2}\trace\left(\Ls \Exp{\sum_{t\in\mathrm{speech}}\scattp{\xvec_{it}-\musi}}\right)\right] =0 \:.
\end{align}

Then,
\begin{align}
    \rsi^{-1} = \frac{1}{d\Nsi}\trace\left(\Ls\Exp{\sum_{t\in\mathrm{speech}}\scattp{\xvec_{it}-\musi}}\right)
\end{align}

Equivalently, for $\rni$,
\begin{align}
    \rni^{-1} = \frac{1}{d\Nni}\trace\left(\Ln\Exp{\sum_{t\in\mathrm{sil}}\scattp{\xvec_{it}-\muni}}\right)
\end{align}

where
\begin{align}
    &\Exp{\sum_{t\in\mathrm{speech}}\scattp{\xvec_{it}-\musi}} = 
    \Ssi - \Fsi\Exp{\musi}^\T - \Exp{\musi} \Fsi^\T + \Nsi\Exp{\musi\musi^\T}\\
    &\Exp{\sum_{t\in\mathrm{sil}}\scattp{\xvec_{it}-\muni}} = 
    \Sni - \Fni\Exp{\muni}^\T - \Exp{\muni}\Fni^\T + \Nni\Exp{\muni\muni^\T}
\end{align}

The above equations have the inconvenience that we cannot compute $\rsi$ (or $\rni$) if no speech (or silence) frames have been observed, since no priors were put on $r$. We propose two possible initializations for $r$: (i) we can choose $r=1$ until some frames are observed, or (ii) a global $r$ pre-computed over the whole training set can be computed as
\begin{align}
    &\rs^{-1} = \frac{1}{d\sum_i \Nsi}\trace\left(\Ls\sum_i\Exp{\sum_{t\in\mathrm{speech}}\scattp{\xvec_{it}-\musi}}\right)\\
    &\rn^{-1} = \frac{1}{d\sum_i \Nni}\trace\left(\Ln\sum_i\Exp{\sum_{t\in\mathrm{sil}}\scattp{\xvec_{it}-\muni}}\right) \;.
\end{align}
\end{document}